\documentstyle[12pt]{article}
\newcommand{\be}{\nopagebreak[3]\begin{equation}}
\newcommand{\ee}{\end{equation}}
\newcommand{\dsp}{\displaystyle}
\newcommand{\txst}{\textstyle}
\newcommand{\ba}[1]{\begin{array}{#1}\dsp}
\newcommand{\ea}{\end{array}}
\newcommand{\nl}{\vspace{1pc} \\ \dsp}
\newcommand{\scs}[1]{{\scriptscriptstyle #1}}
\newcommand{\ovr}[2]{\raisebox{-1ex}{$\stackrel{\dsp #1}{\scs{#2}}$}}
\newcommand{\im}{\mbox{\rm Im}\,}
\newcommand{\re}{\mbox{\rm Re}\,}
\newcommand{\tr}{\mbox{\rm tr}\,}
\newcommand{\eq}[1]{Eq.~(\ref{#1})}

\renewcommand{\d}{\partial}
\newcommand{\dd}[2]{\frac{\partial #1}{\partial #2}}

\newcommand{\N}{\scs N}

\newcommand{\G}{\scs G}
\renewcommand{\a}{\alpha}
\renewcommand{\b}{\beta}
\newcommand{\lam}{\lambda}

\newcommand{\ie}{{\em i.e., }}

\begin{document}
\begin{titlepage}
\begin{flushright}
NBI-HE-94-24\\
April, 1994
\end{flushright}
\vspace*{5pc}
\begin{center}
{\huge \bf
QCD ON A TREE
}\end{center}
\vspace{2pc}
\begin{center}
 {\Large D.V. Boulatov}\\
\vspace{1pc}
{\em The Niels Bohr Institute,\\
University of Copenhagen\\
Blegdamsvej 17, DK-2100 Copenhagen \O, \\
DENMARK}
\vspace{2pc}
\end{center}
\begin{center}
{\large\bf Abstract}
\end{center}
A model is proposed which can be regarded as
a mean field approximation for pure lattice QCD and
chiral field. It always possesses a phase transition between a
strong-coupling phase (where it reduces  to a one-plaquette
integral) and a non-trivial weak-coupling one. For the U$(N)$ gauge group,
it is equivalent to some hermitian multi-matrix model. This analogy
allows for determining possible large $N$ critical regimes thus
generalizing the Gross-Witten  phase transition in the one-plaquette model.
\vfill
\end{titlepage}

\section{Introduction}

The recent few years have seen the considerable development of the
planar-diagram technique connected mostly with the matrix models of 2D
gravity. Unfortunately, it seems that all those achievements has
brought no new insights  into large $N$ gauge theory, which was the
original motivation for the method \cite{plandiag}.
Nevertheless, as was demonstrated by the exact solution of QCD$_2$ on
a sphere \cite{qcd2}, a reduction to a hermitian matrix model can be very
profitable technically.

In the present paper I establish a connection between a special class
of hermitian multi-matrix models  and a mean field approximation for
pure lattice QCD. It enables for using the saddle-point technique for
the former in the rather new framework. The most interesting phenomenon
here is, probably, a large $N$ phase transition of the Gross-Witten
type \cite{GW}, which should have some stringy interpretation.
For the one-plaquette model, the connection with 2D gravity
was discussed in a number of papers \cite{2dgrav}.

The standard QCD mean field (MF), as was proposed by
K.Wilson~\cite{Wilson} (for review see \cite{Drouff}),
 suffers from the obvious drawback of being
gauge dependent.  I suggest a  purely geometrical  approach to the
problem, which avoids the step of gauge fixing. This model,
however, shares all limitations of any MF approximation at the price
of being, in principle, soluble. One can say it learns nothing about
QCD itself. Unfortunately, the same could be said about any other
model solved so far.

The starting point is to substitute a regular D-dimensional lattice  by
an infinite (Cayley) tree constructed  of two-dimensional
plaquettes.
The plaquettes are glued along their edges so that the tree is a simply
connected covering of the lattice.
Therefore, gauge theory defined
on such a tree can be regarded as a MF  approximation  for lattice
QCD in D dimensions.

A similar idea was put forward in Ref.~\cite{Zuber} where gauge theory
on a Cayley tree made of cubes was considered. As far as phase
structures of lattice models are concerned, the cube-made Cayley tree
might provide better accuracy than the plaquette-made one (although it
is not clear a priori). However,
the latter enjoys the property of being soluble in the large $N$
limit by a saddle-point technique, while the former can hardly be handled
for continuous gauge groups. Another nice feature of the model under
consideration is that it includes both chiral field and gauge theory
on equal footing , actually interpolating between them.

The transfer-matrix formalism is very simple in
our case.
Let us consider a tree with one root, \ie to leave one of gauge
variables, $u$, not integrated. Then, the corresponding partition function
$I_V(u)$ for the rooted tree of a large volume $V$ obeys the equation

\be
I_{pmV+1}(u) = \int \prod_{k=1}^p dx_k\; K( u\prod_k^p x_k )
\prod_{k=1}^p I^m_V(x_k)
\ee
where the tree is assumed to be made of $(p+1)$-sided polygons,
$m+1$ on each link. If $p=3$, it is a covering of the
$D=\frac{m+3}2$ dimensional hyper-cubic lattice.
The value $p=1$ corresponds to the case when a plaquette-made tree
degenerates into an ordinary one constructed of one-dimensional links.
Hence, we have a
model interpolating between the spin and gauge MF approximations. For
the $SU(N)$ group, it interpolates between $D_{CF}=\frac{m+1}2$
dimensional chiral field and $D_{GT}=D_{CF}+1$ lattice gauge theory.

 The Boltzmann weight,
$K(x)$, is a real positive class function.  Two standard choices of it
are: the  Wilson one

\be
K(x)=\exp \frac{N}{2g^2}\tr(x+x^+)
\ee
and  the heat-kernel

\be
K(x)=\sum_r d_r e^{-\frac{g^2}N C_r}\chi_r(x)
\ee
where $g^2$ is a bare gauge coupling; $\chi_r(x)$ is a character of an
irrep $r$; $C_r$ is a second Casimir; $d_r=\chi_r(I)$ is the dimension
of $r$.

In the thermodynamical limit, the conveniently normalized quantity

\be
J(u) = \lim_{V\to\infty} e^{-(V+\frac1{mp-1})f} I_V(u)
\ee
(where $f$ is a free energy per volume) obeys the equation

\be
J(u) = \int \prod_{k=1}^p dx_k\; K(u\prod_{k=1}^p x_k)
 \prod_{k=1}^p J^m(x_k)
\label{mfeq}
\ee
and the free energy is given by the relation

\be
f(g^2) = - \frac{pm-1}{m+1} \log \int du\; J^{m+1}(u)
\ee

\eq{mfeq} always has the trivial solution

\be
J_{sc}\equiv j_0 = \Big[ \int dx\; K(x)\Big]^{\frac1{mp-1}}
\ee
In this case, the free energy coincides with the one-plaquette-model one

\be
f_{sc}(g^2)=\log \int dx\; K(x)
\ee
This is the strong-coupling phase of the model.

In the weak-coupling phase, $J(x)$ is a non-trivial class function:

\be
J(x) = \sum_r j_r \chi_r(x)
\label{fexp}
\ee
where the sum runs over all irreps of a gauge
group. \eq{mfeq} can be rewritten in terms of the Fourier coefficients
$j_r$ as

\be
j_r = \lam_r \Big[ \frac1{d_r}\sum_{s_1,\ldots,s_m}
\prod_{k=1}^m j_{s_k} \int dx\; \chi_r(x) \prod_{k=1}^m
\chi_{s_k}(x)\Big]^p
\label{ftmfeq}
\ee
where $\lam_r$ are Fourier coefficients
of the Boltzmann weight.

The free energy can be rewritten as the single sum over representations

\be
f(g^2) = - \frac{pm-1}{m+1} \log \sum_r d_r j_r
\Bigg(\frac{j_r}{\lam_r}\Bigg)^{\frac1p}
\label{fe2}
\ee

A glueball spectrum can be determined from the eigenvalue problem:

\be
e^{-m_G} v_r = \sum_t M_{rt} v_t
\label{eveq}
\ee
where the analog of the transfer matrix in our case is simply

\be\ba{rcl}
M_{rt} &=& \lam_r \Big[\dsp \frac1{d_r}\sum_{s_1,\ldots,s_m}
\prod_{k=1}^m j_{s_k} \int dx\; \chi_r(x) \prod_{k=1}^m
\chi_{s_k}(x)\Big]^{p-1}\nl && \dsp
\frac1{d_r}\sum_{s_1,\ldots,s_{m-1}}
\prod_{k=1}^{m-1} j_{s_k} \int dx\; \chi_r(x) \chi_t(x) \prod_{k=1}^{m-1}
\chi_{s_k}(x)\ea
\label{trmat}
\ee

\eq{eveq} always has the trivial eigenvalue $m_{\G}=0$, which
corresponds to the identity operator (\ie the partition function).
The number of possible excitations in the system is equal to the number
of irreducible representations of a gauge group. In the weak-coupling
phase they are all exited. However, non of them can
become massless  and, hence, there is no continuum limit associated
with  the model.

\section{Phase transition}

Let us assume that, at some critical value $g^2_*$, the Fourier
coefficients of all non-trivial representations in \eq{fexp} vanish (\ie
the weak-coupling solution transforms smoothly into the strong-coupling
one). Then, in the vicinity, $\Delta g^2 = g^2_* - g^2 \ll 1$, the
most essential contribution comes from the fundamental representation, $r=1$,

\be
j_0=O(1);\hspace{2em} j_1=O(\Delta g^2);\hspace{2em} j_r=o(\Delta g^2),\
\mbox{ for}\ r\neq 0,1
\ee
After expanding \eq{ftmfeq} in $j_1$, one finds

\be\ba{l}
j_0 = \lam_0 j_0^{mp} + O(j_1^2)\nl
j_1 = \lam_1[\frac{m}{d_1}j_0^{m-1}j_1]^p + \ldots
\ea
\label{creq}
\ee
{}From which it follows that, for $p\neq1$, the smooth transition
between the strong and weak-coupling solutions is impossible. For $p=1$,
 the critical value $g^2_*$ is determined by the equation

\be
\frac1m=\frac{\lam_1(g^2_*)}{d_1\lam_0(g^2_*)}
\ee
For $SU(N)$ with the Wilson weight, one finds

\be
\frac1m=\frac1N\langle \tr U \rangle \ovr{=}{N\to\infty}\frac1{g^2_*}
\ee
Hence, the phase transition takes place when the one-plaquette model is
in its strong-coupling regime (\ie above the Gross-Witten critical
point). Also notice that, in the $m\to\infty$ limit,
the strong-coupling phase disappears. For the heat-kernel, one finds
simply $g^2_*=\log m$ independent of $N$.

The lowest two representations play a special role in the $p=1$ case.
Therefore, the simplest $Z_2$ model may be rather instructive as
far as the nature of the phase transition is concerned. In this case one
can represent

\be
J(x) = j_0 + j_1x = \rho e^{x\phi}
\ee
where $x=\pm 1$ is a $Z_2$ variable and

\be
j_0=\rho \cosh \phi \hspace{2pc} j_1 = \rho\sinh\phi
\ee
Let me choose $\lam_0=1$, then \eq{ftmfeq} takes the form

\be\ba{l}
\rho\cosh\phi = (\rho^m\cosh m\phi)^p \nl
\rho\sinh\phi = \lam_1(\rho^m\sinh m\phi)^p
\ea
\ee
or, equivalently,

\be
\tanh\phi = \lam_1\tanh^p m\phi \hspace{2pc}
\rho = \bigg(\frac{\cosh\phi}{\cosh^pm\phi}\bigg)^{\frac1{mp-1}}
\label{Z2eq}
\ee
Hence, a solution can always be determined from some algebraic equation
with respect to $x=\tanh\phi$.

For example, for $m=3$ and $p=1$ (2D Ising), one finds the equation

\be
x=\lam_1\frac{3x+x^3}{1+3x^2}
\ee
from which
\be
x=\sqrt{\frac{3\lam_1-1}{3-\lam_1}}\hspace{1pc}
f=\frac12\log
\frac{8\lam_1^2}{6\lam_1-\lam_1^2-1}
\ee
And the transition is of the second order: $f'(\frac13)=0$.
This is the case, if $p=1$, for arbitrary $m$ and all
compact  groups.  In general,
it can be easily proven by using the representation (\ref{fe2}) for
the free energy.

In the $Z_2$ model, there is only one excitation for which, in the
previous example ($m=3,\ p=1$),  one finds the mass
$m_G=\log\frac{4\lam_1}{(1-\lam_1)^2}$.
One can easily show that, at the critical point for all $m$,
$m_{\G}(\lam_1^*)= \log m$.
Therefore, the transition does not produce any continuum limit.

Of course, when $p=1$, one just repeats the standard spin MF
approximation \cite{Bethe}.
If $p\neq1$, the phase transition between the strong and weak-coupling
phases is of the first order. \eq{Z2eq} has obviously no real solutions
for $\lam_1$ small enough. Hence, the weak coupling branch disappears
somewhere being already meta-stable. The first-order transition point
can be determined from the equation

\be
j_0^{\frac{p+1}p}+\lam_1^{-\frac1p}j_1^{\frac{p+1}p}=1
\ee
which is quite a standard  numerical problem.

\section{Connection with multi-matrix models}

The Fourier representation (\ref{ftmfeq}) is not convenient for the
investigation of the weak-coupling phase.
In this case, the original matrix variables are more suitable.
Let us choose the Boltzmann weight in the form of the $U(N)$
heat-kernel:

\be
K(x)=\sum_r d_r e^{-\frac{g^2}{2N}C_r} \chi_r(x)
\label{hker}
\ee
where, in terms of the highest weight components ($m_1\geq m_2\geq
\ldots\geq m_{\N}$),

\be
C_r = \sum_{k=1}^N\Big(m_k-k+N\Big)^2
\ee
is a conveniently redefined second Casimir.
For diagonal matrices, $x=e^{i\a}$, one has the Weyl formula for characters

\be
\chi_r(e^{i\a})=\frac{\Delta_r(e^{i\a})}{\Delta(e^{i\a})}
\ee
where

\be\ba{l}
\Delta_r(e^{i\a}) =
\raisebox{-1.25ex}{$\stackrel{\dsp\det}{\scs{(j,k)}}$}\;
e^{i(m_k-k+N)\a_j}\nl
\Delta(e^{i\a}) =
\raisebox{-1.25ex}{$\stackrel{\dsp\det}{\scs{(j,k)}}$}\;
e^{i(N-k)\a_j}=
\prod_{j<k}(e^{i\a_j}-e^{i\a_k})
\ea\ee

\be
d_r=\chi_r(I)=\prod_{j<k}\Big(1+\frac{m_j-m_k}{k-j}\Big)
\ee
is the dimension of an irrep $r$.

Let us substitute (\ref{hker})  in (\ref{mfeq}) and integrate over
angular parts of all variables:

\[\ba{l}
\int \prod_{i=1}^p dS_i\; K(e^{i\a}\prod_{k=1}^p S_ke^{i\b_k}S_k^+) =
\sum_r d_r^{1-p} e^{-\frac{g^2}{2N}C_r} \chi_r(e^{i\a})\prod_{k=1}^p
\chi_r(e^{i\b_k}) =
\nl
\Big(\prod_{n=1}^{N-1}n!\Big)^{p-1}\frac1{\Delta(e^{i\a})}
\sum_{\ell_1>\ell_2>\ldots>\ell_N} [\Delta(\ell)]^{1-p}
\exp\Big(-\frac{g^2}{2N}\sum_{k=1}^N \ell_k^2\Big)
\raisebox{-1.25ex}{$\stackrel{\dsp\det}{\scs{(j,k)}}$}\;
e^{i\ell_k\a_j}\nl
\prod_{i=1}^p \frac{\det e^{i\ell_j\b_{ik}}}
{\Delta(e^{i\b_i})} =
\frac1{N!}\Big(\prod_{n=1}^{N-1}n!\Big)^{p-1}
\frac1{\Delta(e^{i\a})} \prod_{k=1}^N \Bigg( \sum_{n_k=-\infty}^{+\infty}
\int_{-\infty}^{+\infty} d\lam_k\; \Bigg)
[\Delta(\lam)]^{1-p}
\nl
\exp\Big[-\frac{g^2}{2N}\sum_{k=1}^N \lam_k^2
+i\sum_{k=1}^N \lam_k(\a_k+2\pi n_k)\Big]
\prod_{i=1}^p \frac{\det e^{i\lambda_j\b_{ik}}}
{\Delta(e^{i\b_i})}
\ea\]
The last equality holds owing to Poisson's formula. Now, \eq{mfeq} takes
the form (after renormalizing $J(x)$)

\be\ba{l}
J(e^{i\a}) = \frac1{\Delta(e^{i\a})}
\prod_{k=1}^N \Bigg( \sum_{n_k=-\infty}^{+\infty}
\int_{-\infty}^{+\infty} d\lam_k\; \Bigg)
\Delta(\lam)
\exp\Big[-\frac{g^2N}2\sum_{k=1}^N \lam_k^2  \nl
+iN\sum_{k=1}^N \lam_k(\a_k+2\pi n_k)\Big]
 \Bigg[ \int_0^{2\pi} \prod_{k=1}^N d\b_{k}\; \Delta(e^{i\b})
\frac{ e^{iN\sum\lam_k\b_k}}{\Delta(\lam)}
J^m(e^{i\b})\Bigg]^{p}
\ea\ee

Let us introduce a new function $F(\a)$ such that

\be
\Delta(e^{i\a})J(e^{i\a}) = \sum_{\{n\in Z^N\}} \Delta(\a+2\pi n)
F(\a+2\pi n)
\ee
which obeys the equation

\be\ba{rcl}
F(\a) &=&\dsp \frac1{\Delta(\a)}
\int_{-\infty}^{+\infty} \prod_{k=1}^N d\lam_k\; \Delta(\lam)
\exp\Big[-\frac{g^2N}2\sum_{k=1}^N \lam_k^2
+iN\sum_{k=1}^N \lam_k\a_k\Big] \nl &&\dsp
\Bigg[ \int_{-\infty}^{+\infty} \prod_{k=1}^N d\b_{k}\;
\Delta(\b)
\frac{ e^{iN\sum\lam_k\b_{k}}}{\Delta(\lam)}
J^{m-1}(e^{i\b})F(\b)\Bigg]^p
\ea
\label{mfeqred}
\ee

If one introduces hermitian matrices $A,\ \Lambda,\ B$ having
eigenvalues $\a,\ \lam,\ \b$ correspondingly, then one can rewrite
\eq{mfeqred} as the matrix integral equation

\be\ba{rcl}
F(A) &=&\dsp \int d^{N^2}\Lambda\; e^{-\frac12g^2N\tr \Lambda^2
+iN\tr\Lambda A}  \Big[ \int d^{N^2} B\;
e^{iN\tr\Lambda B +V[B]} F(B)\Big]^p \nl  & = &\dsp
\Bigg(\frac{2\pi}g\Bigg)^{\frac{N^2}2} \int \prod_{i=1}^p d^{N^2} B_i\;
e^{-\frac{N}{2g^2}\tr(A+\sum B_i)^2 + \sum V[B_i]}\prod_{i=1}^p F(B_i)
\ea
\label{matrep}
\ee
where the effective potential is determined by the equation

\be
V[B] = (m-1)\log\sum_{\{n\in Z^N\}} \frac{\Delta(\b+2\pi n)}
{\Delta(e^{i\b})} F(\b+2\pi n)
\label{efpot}
\ee
$V[B]$ is a symmetric non-singular function of eigenvalues, hence,
can be expanded in Schur polynomials.
It is easily checked that, if $m=1$, $F(\b)$ is Gaussian and
$J(e^{i\b})$ is given by the heat-kernel in the Dowker form \cite{Dowker}.

\section{Large $N$ solution}

If $p=1$, nothing prevents one from introducing an arbitrary potential,
$N\tr U(B)$, for the chiral field. In this case, what one ends up with is
a unitary analog of the Bethe-tree matrix model considered in
Ref.~\cite{Btmm}. One can use a similar technique in both cases. The first
step is to write down the self-consistency equation for the resolvent
matrix  \cite{Boul}:

\be\ba{l}
F(A) = \Bigg(\frac{2\pi}g\Bigg)^{\frac{N^2}2} \int d^{N^2} B\;
e^{-\frac{N}{2g^2}\tr(A+B)^2 + V[B]-N\tr U(B)} F(B) =
\Bigg(\frac{2\pi}{g}\Bigg)^{\frac{N^2}2} \nl  \int d^{N^2} B\;
\frac1N \tr \Bigg[\Big(z+A+\frac{g^2}N\frac{\d\ }{\d A}\Big)
\frac1{z-B}\Bigg]
e^{-\frac{N}{2g^2}\tr(A+B)^2 + V[B]-N\tr U(B)} F(B)  \nl
=\frac1N \sum_{k=1}^N \Bigg\{(z+\frac{g^2}N\frac{\d\ }{\d \a_k}+\a_k)G_k(z) +
\frac{g^2}N \sum_{j\neq k}\frac{G_k(z)-G_j(z)}{\a_k-\a_j}\Bigg\}
\ea
\label{SDeq}
\ee
where

\be
G_k(z) = \Big(\frac{2\pi}g\Big)^{\frac{N^2}{2}}
\int dB\; \Bigg[\frac1{z-B}\Bigg]_{kk}
e^{-\frac{N}{2g^2}\tr(A+B)^2 + V[B]-N\tr U(B)} F(B)
\label{Gk=}
\ee
are diagonal matrix elements of the resolvent matrix. This equation is
a recursive relation for moments of $B$.

Let us introduce two functions

\be\ba{l}
f(x) = \langle\frac1N \tr\frac1{x-A}\rangle \ovr{=}{N\to\infty}
\int dy \frac{\rho(y)}{x-y} \nl
F(x,z) = \langle\frac1N \tr\frac1{x-A}\frac1{z-B}\rangle
\ovr{=}{N\to\infty} \int dy \frac{\rho(y)W(y,z)}{x-y}
\ea
\label{fF}
\ee
$\rho(y)$ is a density of eigenvalues; $W(y,z)$ is a real function on
a support of $\rho(y)$. As we are looking for a
homogeneous ground state, $f(x)$ is the same at all sites of the tree
and $F(x,z)$ is symmetric: $F(x,z)=F(z,x)$.

{}From \eq{SDeq} it follows that, at $N=\infty$,  $W(x,z)$ obeys the
equation

\be
(z+g^2w(x)+x)W(x,z)+g^2\int dy\; \rho(y) \frac{W(x,z)-W(y,z)}{x-y}=1
\label{eqforW}
\ee
where

\be
w(x)= \lim_{N\to\infty}\frac1{N}\dd{\ }{\a_k}\log F[A]\Big|_{\a_k=x}
=\dd{\ }{x}\frac{\delta\ }{\delta \rho(x)}
\lim_{N\to\infty}\frac1{N^2}\log F
\ee

As was first noticed in Ref.~\cite{Mig},
equations of this type can be solved by
the Riemann-Hilbert method.  The outcome of which is the following integral
representation

\be
F(x,z) = 1-\exp \int \frac{dy}{2\pi i} \frac1{x-y}\log
\frac{z-u_+(y)}{z-u_-(y)}
\label{F=}
\ee
where, in our case,

\be
u_{\pm}(x) =-x-g^2( w(x)+\re f(x) \pm i\; \im f(x))
\label{upm=}
\ee
and the integral goes along a support of $\im f(y)=\pi\rho(y)$.
Off the support of $\rho(x)$, $\re f(x)$ and $\im f(x)$ continue
analytically as two independent holomorphic functions. \eq{F=} makes
sense as a set of recursive integral relations obtained by expanding
both sides in inverse powers of $x$ and $z$  (see
Ref.~\cite{Btmm} for details).

As $F(x,z)$ is symmetric, the following equation holds

\be
u_+(u_-(x))=x
\label{uu=x}
\ee

The function $w(x)$ can be determined from the saddle-point equation

\be
2\re f(x) + 2w(x) + \dd{\ }{x}\frac{\delta\ }{\delta \rho(x)}
\lim_{N\to\infty}\frac1{N^2} V[x] -U'(x) = 0
\label{speq}
\ee
which is, in general, a  complicated non-linear integral equation.
However, for $g^2$ small enough, we can neglect
in the large $N$ limit all non-trivial
winding numbers in \eq{efpot} (\ie put $n_k=0\ \forall k$), then \eq{speq}
takes the simple form

\be
2\re f(x) + (m+1)w(x) - (m-1)\int dy\; \rho(y)\Bigg(\frac12\cot\frac{x-y}2
-\frac1{x-y}\Bigg) - U'(x)= 0
\ee
or

\be
w(x) = \frac1{m+1}\Bigg\{U'(x)-2\re f(x) + (m-1)
\sum_{n=-\infty}^{+\infty}\!\!\!\!\mbox{}'\Big[
 f(x-2\pi n) + \frac1{2\pi n}
\Big]\Bigg\}
\ee
The prime means that the $n=0$ term in the sum is omitted.
This formula makes sense only when a support of $\rho(x)$ lies inside
the interval $(-\pi,+\pi)$, \ie when there is
a gap in the eigenvalue distribution of original unitary matrices.

If $p>1$, any local potential for unitary variables spoils the gauge
invariance of the model. However, one can introduce a non-gaussian
potential for the auxiliary  field $\Lambda$ in \eq{matrep}. It
corresponds to taking an arbitrary Boltzmann weight. In this case, one
cannot easily integrate out the auxiliary field.  Nevertheless, our
method can be generalized to the inhomogeneous system. Now, one has
two different external field problems and has to introduce three functions

\be\ba{l}
f(x) = \langle\frac1N \tr\frac1{x-B}\rangle \ovr{=}{N\to\infty}
\int dy \frac{\rho(y)}{x-y} \nl
\varphi(z) = \langle\frac1N \tr\frac1{z-\Lambda}\rangle \ovr{=}{N\to\infty}
\int d\lam \frac{\eta(\lam)}{z-\lam} \nl
F(x,z) = \langle\frac1N \tr\frac1{x-B}\frac1{z-\Lambda}\rangle
\ovr{=}{N\to\infty} \int dy \frac{\rho(y)W(y,z)}{x-y}  =
\int d\lam \frac{\eta(\lam)\Omega(\lam,x)}{z-\lam}
\ea
\label{fphiF}
\ee
Both $W(x,z)$ and $\Omega(\lam,x)$ obey equations analogous to
(\ref{eqforW}) and $F(x,z)$ has two different integral representations
of the type (\ref{F=}). For example,

\be\ba{l}
I(\lambda) = \int dB\; e^{iN\tr \Lambda B + V[B]} F(B) = \nl
\int dB \frac1N \tr\Bigg[\Big(x+\frac{i}N\dd{\ }{\Lambda}\Big)
\frac1{x-B}\Bigg] e^{iN\tr \Lambda B+V[B]}F(B) = \nl
\frac1N\sum_{k=1}^N\Bigg\{\Big(x+\frac{i}N\dd{\ }{\lam_k}\Big)G_k(x)
+\frac{i}N\sum_{j\neq k} \frac{G_k(x)-G_j(x)}{\lam_k-\lam_j}\Bigg\}
\ea\ee
where $G_k(x)$ are diagonal elements of the resolvent matrix as in
\eq{Gk=}. Then one finds

\be
(x+i\omega(\lam))\Omega(\lam,x) + i \int d\mu\; \eta(\mu)
\frac{\Omega(\lam,x)-\Omega(\mu,x)}{\lam-\mu}=1
\ee
where

\be
\omega(\lam)=\lim_{N\to\infty}\frac1N\dd{\ }{\lam_k} \log I(\Lambda)
\Big|_{\lam_k=\lam}
= \dd{\ }{\lam}\frac{\delta}{\delta \eta(\lam)}
\lim_{N\to\infty}\frac1{N^2} I(\Lambda)
\ee

The saddle-point equation
with respect to $\Lambda$ gives

\be
2\re \varphi(\lam) - U'(\lam) +(p+1)\omega(\lam)=0
\ee
where $U(\lam)$ is an arbitrary, in principle, potential.

Thus, one finds for $F(x,z)$ the representation

\be
F(x,z)=1-\exp\int\frac{d\lam}{2\pi i}\frac1{z-\lam}
\log\frac{x-v_+(\lam)}{x-v_-(\lam)}
\ee
where

\be
v_{\pm}(\lam)=\frac1i\Big\{\frac1{p+1}U'(\lam) +
\frac{p-1}{p+1}\re\varphi(\lam) \pm i \im\varphi(\lam)\Big\}
\ee
In a close analogy, one finds

\be
(z+iw(x))W(x,z) + i \int dy\; \rho(y)
\frac{W(x,z)-W(y,z)}{x-y}=1
\ee
and the saddle-point equation for a distribution with a gap:

\be
2\re f(x) + (m+1)w(x) - (m-1){\sum_{n=-\infty}^{+\infty}}\!\!\mbox{}'
\Big[{\txst f(x-2\pi n) + \frac1{2\pi n}}\Big]=0
\ee
Hence,

\be
F(x,z) = 1-\exp\int \frac{dy}{2\pi i}
\frac1{x-y}\log\frac{z-u_+(y)}{z-u_-(y)}
\ee
where

\be
u_{\pm}(x) = \frac1i\Big\{\frac{m-1}{m+1}\re f(x) +
\frac{m-1}{m+1}{\sum_{n=-\infty}^{+\infty}}\!\!\mbox{}'
\Big[f(x-2\pi n) + \frac1{2\pi n}\Big] \pm i\im f(x)
\Big\}
\ee
And, instead of \eq{uu=x}, there are two equations

\be
u_+(v_-(x))=x \hspace{2pc} u_-(v_+(x))=x
\label{uv=x}
\ee

\section{Critical   regimes}

Following Ref.~\cite{Btmm}, one can determine possible large $N$
critical regimes. Eqs.~(\ref{uu=x}) and (\ref{uv=x}) allows, in
principle, for constructing  the functions $f(x)$ and $\varphi(x)$
thus solving  the model at large $N$. Practically, it is a very
complicated problem. However, universal behavior is determined by a
scaling of the imaginary parts of $f(x)$ and $\varphi(x)$ near their
edges, which are, in general, branching points:

\be
f(x)=f_{reg}(x) + c (x-x_0)^{1+\gamma}\ldots \hspace{1.5em}
\varphi(\lam)=\varphi_{reg}(\lam)+c'(\lam-\lam_0)^{1+\gamma}\ldots
\ee
$f_{\rm reg}(x)$ and $\phi_{\rm reg}(x)$ are regular parts of the
functions at the branching points.

Let us start with the $p=1$ model.
We are interested in the situation when the edges of the distribution of
eigenvalues of original unitary matrices collide. It
corresponds to the case when the edges of $\im f(x)$ in \eq{upm=} touch
branching points of $f(x\pm2\pi)$. Let us expand all
quantities in \eq{upm=} near one of the collision points (by
redefining the variables one can always place it at the origin):

\be
u_{\pm}(x) = ax+b\frac{m-1}{m+1}(\cos\pi\gamma+1)
e^{-i\pi\gamma}x^{1+\gamma} \pm ib\sin\pi\gamma e^{-i\pi\gamma}x^{1+\gamma}
+\ldots
\ee
By substituting it in \eq{uu=x}, one obtains two equations

\be
a^2=1
\ee
and

\be
\frac{m-1}{m+1}(\cos\pi\gamma+1) - i\sin\pi\gamma +
a^{\gamma}\Big[\frac{m-1}{m+1}(\cos\pi\gamma+1) + i\sin\pi\gamma\Big]=0
\ee
The simplest possibility is $a=1$, when one finds

\be
\cos\pi\gamma=-1
\label{gw}
\ee
which is the standard Gross-Witten singularity always possible in the
model. By tuning the potential $U(x)$, one can reach a multi-critical
point when \mbox{$a=-1$}, then there are two branches:
(i)~$a^{\gamma}=e^{i\pi\gamma}$ yielding the equation for $\gamma$

\be
\cos\pi\gamma =\frac1m
\label{m>1}
\ee
and (ii)~$a^{\gamma}=e^{-i\pi\gamma}$ yielding

\be
\cos\pi\gamma = m
\ee

Let me remind a reader that $m$ is connected with the effective space
dimension
as $m=2D-1$. The obvious duality $m\to\frac1m$ is just a manifestation
of the duality $D\to\frac{D}{2D-1}$ in the Bethe-tree matrix model
found in Ref.~\cite{Btmm}. The solution fits two soluble cases:
$m=-1$, which corresponds to the one-plaquette model, and m=1, which
is just a one-dimensional matrix chain. The case $m=0$ corresponds to
a two-matrix model (as we have introduced an arbitrary potential, it
is not simply reducible to a one-matrix integral any more).

If $p>1$, there are two functions obeying \eq{uv=x}.
Expanding  them as

\be\ba{l}
u_{\pm}(x) = ax-b\frac{m-1}{m+1}(\cos\pi\gamma+1)
e^{-i\pi\gamma}x^{1+\gamma} \pm ib\sin\pi\gamma e^{-i\pi\gamma}x^{1+\gamma}
+\ldots \nl
v_{\pm}(x) = a'x-b'\frac{p-1}{p+1}\cos\pi\gamma
e^{-i\pi\gamma}x^{1+\gamma} \pm ib'\sin\pi\gamma e^{-i\pi\gamma}x^{1+\gamma}
+\ldots
\ea\ee
and substituting the expansions in Eqs.~(\ref{uv=x}), one finds that
$aa'=1$ and

\be
a'b\Big[\frac{m-1}{m+1}(\cos\pi\gamma+1)\pm i\sin\pi\gamma\Big] -
b'a^{1+\gamma}\Big[-\frac{p-1}{p+1}\cos\pi\gamma\pm i\sin\pi\gamma\Big]=0
\ee
This is a degenerate system of linear equations for the real and
imaginary parts. Equating to zero the corresponding determinant, one finds

\be
\cos\pi\gamma = -\frac1{1+\frac{p-1}{p+1}\frac{m+1}{m-1}}
\label{gtscal}
\ee
At $p=1$, this formula reproduces \eq{gw} and, at $m=1$, gives
$\gamma=\frac12$, which is the simplest allowed singularity in a
general matrix model.

The critical regime (\ref{gtscal}) corresponds to the situation when
the potential for the $\Lambda$ variable becomes critical (in the standard
matrix-model sense) at the same moment as the edges of the density for
the unitary matrices  collide.  It is a multi-critical point with
respect to the Gross-Witten singularity, which is always allowed and
corresponds to the case where $u_{+}(x)$ has a quadratic
extremum matching with a square-root branching point of $v_{-}(x)$.

\section{Discussion}

Our method allows for determining possible large $N$ critical regimes
without really solving a model. It means that some critical points can
correspond to  non-stable or non-unitary models. It was indeed the
case for the Bethe-tree matrix model at $D>1$, where critical
potentials appeared to be, in general, complex \cite{Btmm}. As our MF
approximation for chiral field is quite similar to the one for
hermitian field, one can expect that the $m>1$ branch (\ref{m>1})
for the $p=1$
model exists only owing to the formal duality, in the sense of
analytical continuation into an unphysical region of parameters. On
the other hand, the scaling (\ref{gtscal}) in the gauge MF model
should be quite sensible for $p>1$ and $m>1$ as corresponds to the
simplest reachable singularity. However, to determine a critical
form of the Boltzmann weight is the crux of our approach.

Usually, the Gross-Witten transition is considered as a pure lattice
artifact having no physical meaning. However, it is a very general
phenomenon taking place in any model reducible to a saddle-point
problem for a unitary-matrix-valued master field.

\section{Acknowledgments}

I thank J.Ambj\o rn, V.A.Kazakov and B.E.Rusakov for helpful
discussions and the CERN theory division, where this work has been
finished, for warm hospitality.


\end{document}